\documentclass[aps,twocolumn,amsmath,amssymb,preprintnumbers]{revtex4}

\usepackage{dcolumn}
\usepackage{bm}
\usepackage{amsfonts,latexsym,graphicx}
\usepackage[applemac]{inputenc}
\usepackage{color}
\usepackage{multirow}

\begin{document}
\title{Characterization of the degree of Musical non-Markovianity}

\author{Maria Mannone$^{1}$, Giuseppe Compagno$^{2}$}
\affiliation{
$^1$School of Music,
University of Minnesota, 100 Ferguson Hall, 2106 4th Street South, Minneapolis, MN 55455, USA\\$^2$Dipartimento di Fisica, Università di Palermo, via Archirafi 36, 90123 Palermo, Italy}

\email{manno012@umn.edu}

\begin{abstract}

Recently, as an aid for musical analysis, in computational musicology mathematical and informatics tools have been developed to quantitatively characterize some aspects of musical compositions.
To a musical composition can be attributed by ear a certain amount of memory. This results associated to repetitions and similarities of the patterns in musical scores.
To higher variations, a lower amount of memory is perceived.  However musical memory of a score has never  been quantitatively defined. Here we aim to give such a measure following an approach similar to that used in Physics to quantify the memory (non-Markovianity) of open quantum systems.
We have applied this measure to some existing musical compositions, showing that the results obtained via this quantifier agree with what one expects by ear.
The Musical non-Markovianity quantifier can thus be used as a new tool that can aid quantitative musical analysis.

\end{abstract}

\maketitle

\section{Introduction \label{intro}}

Recently in computational musicology methods to mathematically quantify specific aspects of musical compositions have been developed \cite{Bel}. 
They are used both as an aid for musical analysis, and to investigate some aspects of samples of compositions \cite{Mazzola, topos}. In particular, these have been applied to analyze: musical theory in a geometrical way \cite{Andreatta}, gesture in music \cite{mazzola_gesture}, counterpoint, interpretation, harmony \cite{topos}, common patterns in sounds amplitude \cite{Mendes} and chords succession \cite{Tymoczko}. Algorithms have also been developed to find the most repeated sequences \cite{Meredith, Meredith2, utgoff}, to compose music \cite{OpenMusic1, Collins_Laney, Collins}, to orchestrate \cite{mannone_IRCAM} and to improvise in real time \cite{assayag, cont, Noll}.
Repetition in music is important, and algorithms have also been used, either to generate music or to detect repeated sequences, \textit{to characterize sequences as something definite} \cite{Schenker, Tempereley, lerdhal, sciarrino, reti, Kramer}.
In particular, Markovian chains and hidden Markov chains \cite{hidden, Flexer, amiot} have been used in the techniques of score following \cite{assayag, cont}, to analyzing chord sequences in jazz improvisation \cite{franz}, to study spectral similarity \cite{Flexer}, and to compose Markovian stochastic music \cite{xenakis}.

Memory is also an important concept in the physics of open systems, that is of systems embedded in an environment. It is associated to the concept of non-Markovianity in the dynamics of the system as distinct from the concept of Markovianity that characterizes a memoryless evolution \cite{breuer_petruccione}. Therefore we shall use the term non-Markovian music to characterize the presence of memory in a musical score.
While quantification of the degree of memory in open systems is still an open problem, recently various criteria have been developed to quantify it \cite{andersson, breuer2009, plenio}. 
Recently the relationship among various criteria has started to be studied in detail \cite{mannone_compagno}.

Standard music characteristically presents repetitions of sequences.
The amount of these repetitions can be in some way considered to be related to the degree of memory of the composition.

Here we shall address the question: is it possible to quantify the degree of memory in existing compositions? In fact, while algorithms to find the theme have been developed, quantification of the degree of memory in a musical composition, at the best of our knowledge, has never been addressed.

The aim of this paper is to develop a method to define and measure the Musical non-Markovianity degree of musical compositions, adopting concepts and mathematical techniques used in Quantum Physics. Among the various criteria to measure the degree of memory, the one that will be applied to the case of musical composition is the one that utilizes the distance between matrices \cite{breuer2009}.

In order to define quantitatively the degree of memory in a musical composition, we follow the way in which the degree of memory is utilized, in the context of the theory of Open Quantum Systems \cite{cover, preskill}.
It is, however, useful to remark that the acceptation of \textit{memory}, and then the definition of non-Markovianity conceived expressly for music, must be different to the definition considered in the quantum case, where the comparisons are made between different states at the same instant. However, in a musical composition we compare sequences of finite duration (starting in different time instants) of the same musical piece. The common aspect is the idea of memory as conservation of characteristics which make a pattern distinguishable from another, like a state from another.	
	
As an application, it will be quantified the memory degree of three different compositions, already by ear appear to present a very different degree of memory, to see if the results obtained agree well with the listeners' expectations.
	
The structure of this paper is as follows.
In chapter \ref{entropy} we briefly discuss the use of entropy and then, in \ref{criteria}, we introduce non-Markovianity criteria used in Quantum Physics, giving more details about which ones are conceptually applicable to musical cases, and in which way. In chapter \ref{musical non-Markovianity} we define musical non-Markovianity. In chapter \ref{realization} we give technical details about the musical matrices defined and the algorithm developed to find them and then we apply our method to fictitious examples. In chapter \ref{examples}, we apply the same method to existing compositions, giving numerical results. In chapter \ref{conclusions} we give some conclusions; in the appendix there is an example of distribution matrices.

\maketitle

\section{Mathematical tools}\label{tools}

\subsection{Relative entropy as a measure of memory \label{entropy}}

In music, repetitions of some meaningful unit is an important feature.
The unit 'theme' can be taken as a such meaningful unit. It usually appears at the beginning of the composition with characteristics that are repeated during the piece. When a clear 'theme' cannot be recognized it is yet possible to individuate an 'initial musical idea'.
The repetitions of recognizable units can be taken as an indication of the 'memory' embedded in the composition.
To be able to quantify some regularity in the score, one must refer to measurable parameters relative to the score, such as pitches, durations, intensities.

Another way to analyze the regularity of the structure in a musical composition, is the measure of the randomness. These two aspects, the amount of repetitions of some musical pattern, and the degree of randomness in the entire composition, can be useful to characterize the musical score.

To quantify the amount of randomness, entropy can be used \cite{preskill}.
The entropy $H(X)$ of a string of N elements of a random variable X, with the measurable parameter x, is defined as:
\begin{equation}
H_{P_I}(X)=-\sum_{x=1}^N p(x_i)\log_2p(x_i),
\end{equation}
where $p(x)$ is the probability that the quantity $X$ has the value $x_i$ for the i-th element of the string, $H(X)$ upper limit being $H_{max}=\ln_2N$ and its lower limit zero. The magnitude of $H(X)$ is taken to quantify the amount of randomness in the musical sequence.

Entropy has in fact been used to quantify the amount of randomness of sequences of pitches in a musical score \cite{xenakis}. In that case, given a string of N notes as elements, the measurable parameter x has been identified with values that are related to the value of the pitch.

However, the entropy is blind respect to the order of appearing of pitches in the sequence. The order is an aspect very related to the memory, and to the recognizability of a melodic line. The use of entropy is then not adequate to characterize well the amount of memory in a musical score. 

To evaluate the memory, that is a measure of repetitions of sequences in a chosen musical score, it is then necessary to use other quantifiers. The \textit{relative entropy} could be then more useful in this context. One can argue that two musical compositions with similar values of relative entropy can have a similar structure.
Relative entropy is obtained by comparing the probability of two different distributions of identical objects. If there are two strings, their relative entropy $D(X)$  (Kullback-Leibler divergence) is defined as \cite{preskill}:
\begin{equation}\label{kullback}
D(X)=-\sum_{x_i=1}^Np(x_i)\log_2\frac{p(x_i)}{q(x_i)},
\end{equation}
where $p(x)$ is the probability that the quantity $X$ has the $x_i$ value for the i-th element of the first string, and $q(x)$ is the corresponding probability for the second string.

The Kullback-Kleiber divergence in its symmetrized form has been used to evaluate the similarity within a musical composition \cite{cont}, giving a measure of the most repeated musical sequences.

When applied to musical strings of pitches, also the use of $D(x)$ is beset with some difficulties. First, $D(X)$ isn't commutative; second, it is required that the two sequences contain the same number of symbols.
The non-commutativity implies that probabilities do not appear in a symmetric way depending on which order is utilized: $p|q$ is different from $q|p$.
As example, the comparison between a musical sequence containing notes C and D and a sequence containing also E that uses relative entropy, requires to neglect the E in the second sequence.
If we eventually consider some rests at the place of missing notes in the shorter string, we obtain a theme that is different from the original one, because rests become a characterizing part of the theme.

Thus it remains open the question if it is possible to develop a measure of \textit{memory}, or the degree of Musical non-Markovianity, without being subject to the restriction limited to the use of relative entropy.

Finally, musical memory is the memory of all elements in the score. 
In order to completely characterize the score content, we need a more complete analysis, that requires the use of several parameters, like intensities and durations.

\subsection{Non-Markovianity criteria \label{criteria}}

The aim of this paragraph is to give a quantitative characterization of non-Markovianity in music. We shall build a measure following a similar path as that used in the theory of Open Quantum Systems (OQS).

Account must however be taken that \textit{memory} in music can be different from what is used in the physics of OQS.
In OQS loss of memory occurs when two states (represented by matrices), initially distinct, become progressively indistinguishable; i.e. the lower the distinguishability, the lower the memory.
 
In a musical score memory is associated to repetitions of patterns. To higher number of repetitions, correspond higher amount of memory, i.e. the lower the distinguishability among sequences, the higher the memory.

Quantification of non-Markovianity in the theory of OQS had been addressed recently by \cite{breuer_petruccione}, introducing different criteria with appropriated quantifiers.
A given quantifier may result in being simpler than another.
Among the quantifiers used in the OQS, one \cite{breuer2009} quantifies memory by considering the variation of the distances of two quantum states with time. The increase or decrease of this quantifier is associated to persistence or loss of memory.

Other criteria study the density matrix that describes the state (the master equation) \cite{preskill}. One criterion looks at the separability of the map: when it is separable, the dynamics is Markovian; else, the dynamics is non-Markovian \cite{plenio}.
One other criterion looks at the signs of coefficients in the master equation \cite{andersson}.
In general, the different criteria of non-Markovianity do not always agree. For example to describe the presence of memory revivals in different time regions \cite{chruscinski}, the comparison among them is an open problem \cite{mannone_compagno}.

We choose to quantify the presence of memory in musical scores a criterion that makes a direct use of matrices.
This because, for a musical score, we do not have anything corresponding to a map or a master equation, while it is instead possible to represent a musical sequence as a matrix.
The distance between two states is given by the trace distance of the matrices $\rho_1$ and $\rho_2$ representing the states \cite{breuer2009}:
\begin{equation}\label{distance}
D(\rho_1,\rho_2)=\frac{1}{2}tr|\rho_1-\rho_2|,
\end{equation}
where, if A is an Hermitian matrix, $|A|=\sqrt{A^\dag A}$.
The rate of variation of the distance $D$ is defined as
\begin{equation}\label{sigma}
\sigma=\sigma(\rho_1,\rho_2)=\frac{d}{dt}D(\rho_1,\rho_2).
\end{equation}
When $\sigma>0$ the distance is increasing, thus the distinguishability between states is preserved. When $\sigma\leq0$ the distance decreases and the memory is progressively lost.

\maketitle

\section{Musical matrices}\label{realization}

In the Physics of OQS, matrices $\rho_1, \rho_2$ represent quantum states; here, matrices must be related to the content of a musical composition.

To adopt similar methods, also in musical compositions, to quantify quantify the amount of frequency, we shall build matrices starting from a musical score. In order to construct these matrices, we shall convert into numbers the symbolic information contained in a musical score.

A musical composition is divisible into sequences, each sequence made to correspond to a set of numerical parameters relative to frequencies, intensities, times of start and durations. Matrices can be constructed in term of these numerical values. The matrices we use represent the distribution relative to every note in each sequence around the mean value of these parameters.

For the heights of notes, we use differences in Hz. In particular we have chosen differences of semitones respect to the middle C.
For the intensities, we use dimensionless numbers to indicate relative intensity indications in musical scores, in particular we choose 90 for $\mathbf{ffff}$, 80 for $\mathbf{fff}$, 70 for $\mathbf{ff}$, 60 for $\mathbf{f}$, 50 for $\mathbf{mf}$, 40 for $\mathbf{mp}$, 30 for $\mathbf{p}$, and so on.
For the times, we have used dimensionless units: that is the ratio between the duration of the examined note and the metronomic unity.
In particular, we have taken 1 for semiquavers, 2 for quavers, and so on.
The duration of a rest will be counted as the time before the start of the following note.
The couples of parameters that we have chosen to characterize our matrices are: duration-intensity, frequency-start, start-intensity, frequency-duration and frequency-intensity.

For each musical sequence, we have first obtained the mean value relative to each parameter. Then, we have evaluated the distance (normalized between 0 and 1) of each parameter of a note in the sequence respect to its mean values.
The range of distances from the first parameter and the range of distance from the second one has been divided into equal parts.

Fixed a sequence, let us consider a note and a couple of its parameter, for example intensity and duration. We evaluate the difference between the value of each parameter of the note and the correspondent mean value in the sequence. We then discretize the bidimensional domain of values, defining a reticular step.
Given a note and its parameters, we look if the normalized difference is contained in each element of the lattice. The number of elements in the lattice represents the musical matrix.

For each couple of parameters one construct a matrix relative to a numerical sequence
It is clear that it is possible to divide the distance range 0-1 into several parts. In our analysis we shall limit to matrices with four rows and four columns. An example of this procedure in the case of the frequency-start matrix, corresponding to a given musical sequence is described in the Appendix. This procedure can be applied through the use of an algorithm that we have developed.

\maketitle

\section{Musical non-Markovianity}\label{musical non-Markovianity}

In the following we shall analyze the structure of a generic musical composition that will be used as basis to define \textit{musical memory}.

Most musical composition are divisible into sections.
The structure of the entire musical piece is determined by the structure of sections.

The most common form, in classical compositions, is A - B - A'. Let us consider a simplified structure A - B. Every section contains several sequences; let us suppose $A_1$, $A_2$, $A_3$ for the section A, and $B_1$, $B_2$, $B_3$ for the section B.
The natural succession in time of the sequences in a score is thus the following: 
\begin{equation*}
\xrightarrow{\rm A_1\,A_2\,A_3\,B_1\,B_2\,B_3}
t
\end{equation*}
In order to define a non-Markovianity criterion, we  reorder these sequences in time to compare them.
Sections A and B are ordered as: 

\begin{equation*}
t
\\ \Big\downarrow
\begin{matrix}
A_1 & B_1
\\ A_2 & B_2
\\ A_3 & B_3
\end{matrix}
\end{equation*}
with time flowing downward.

To each sequence, we assign a set of two-dimensional matrices, each matrix representing the distribution of a couple of variables for each note (for example frequency and duration).
The exact procedure to build these matrices has been given in chapter \ref{realization}.
Here we proceed to apply the criterion of non-Markovianity before described (chapter \ref{criteria}) to the above canonical structure, constructing a matrix that corresponds to a musical sequence.


To the musical composition above correspond the set of matrices:
 \begin{equation*}
t
\\ \Big\downarrow
\begin{matrix}
\rho_1^A & \rho_1^B
\\ \rho_2^A & \rho_2^B
\\ \rho_3^A & \rho_3^B
\end{matrix}
\end{equation*}
The trace distance defined in eq. \ref{distance} between \textit{simultaneous} sequences $\rho_i^A$, $\rho_i^B$ is $D_i=D_i(\rho_i^A,\rho_i^B)$. It measures the distance between the sequences $\rho_i^A$ and $\rho_i^B$.
The variation rate $\sigma$ in the musical case can be defined as $\frac{\Delta D}{\Delta t}$ (we are considering finite time intervals), where $\Delta D_{i+1}=D_i(\rho_{i+1}^A,\rho_{i+1}^B) - D_i(\rho_i^A,\rho_i^B)$ and $\Delta t_{i+1} = t_{start,\,i+1}-t_{start,\,i}$. 

Because of the normalization on the number of notes, all time intervals $\Delta t$ become equal, and we simply define the rates as $\sigma_i=D_{i+1}-D_i$. We consider only positive values, which represent the case of increasing distance.

Contrarily to the case of physical systems where if the distance decrease memory is lost, in a musical composition memory is preserved if the distance is constant or decreases. In physics, the lower the distinguishability, the lower the memory, while in music, the lower the distinguishability, the higher the memory.
In fact, to successive equal musical sequences, that means a maximum musical memory, correspond equal matrices.

Now we define a quantitative measure of the amount of memory, or of non-Markovianity, in a musical composition. For open quantum systems the degree $\mathcal{N}_{max}$ of non-Markovianity is defined as an integration of $\sigma(\rho_1,\rho_2)$ (eq. \ref{sigma}) over time. The contribution to the integral is taken only over regions with $\sigma(\rho_1,\rho_2)>0$, and a maximization is then performed over all the possible initial states.
In the musical case, however, we want to define the memory degree for each singular musical piece and thus for a single initial state, not for all possible different initial states.

We can partially preserve the idea of integration used in physics. Since in a generic musical composition there is not a statistical structure, we can utilize a discretized version of $\mathcal{N}_{max}$, with the sum of positive rates $\sigma_i$ instead of the integral, where no maximization is performed: we call it $n=\sum_{i(if\,\sigma_i>0)}\sigma_i$.

Then, we renormalize this quantity in order to define an $\mathcal{N}$ comprised between 0 and 1: $\mathcal{N}=\frac{\sum_{i(if\,\sigma_i>0)}\sigma_i}{1+\sum_{i(if\,\sigma_i>0)}\sigma_i}$, with the form $\frac{n}{1+n}$ \cite{plenio}. Thus $0\leq\mathcal{N}\leq1$.

Finally, we define our quantifier of Musical non-Markovianity as $\mathcal{M}=1-\mathcal{N}$, to characterize the amount of memory in musical compositions.
The quantifier $\mathcal{M}$ is equal to 1 when the thematic memory is the maximum, i.e. when in a musical composition there are only repetitions of the same sequence. To lower $\mathcal{M}$, it corresponds a lower amount of musical memory. 

The quantifier $\mathcal{M}$ is then
\begin{equation}\label{our quantifier without correction}
\begin{split}
\mathcal{M}=&1-\frac{\sum_{i\,\left(if\,\sigma_i>0\right)}\sigma_i}{1+\sum_{i\,\left(if\,\sigma_i>0\right)}\sigma_i}=
\\
=& \frac{1}{1+\sum_{i\,\left(if\,\sigma_i>0\right)}\sigma_i}.
\end{split}
\end{equation}
$\mathcal{M}$ as defined in eq. \ref{our quantifier without correction} does not differentiate cases with an identical sum of positive rates $\sum_{i,\,\sigma_i>0}\sigma_i$, because it does not give any information about the total number of rates, positive and negative and null.
For example, let us consider a score with ten total rates, with only two positive ones, with sum $\sum_{i,\,\sigma_i>0}\sigma_i^\ast$: the positive contribution has the proportion of 2 over 10. Let us consider another composition, with only three total rates with two positive ones, with identical sum $\sum_{i,\,\sigma_i>0}\sigma_i^\ast$: the positive contribution in this case is 2 over 3. The information given by eq. \ref{our quantifier without correction} is the same for the two compositions, and does not take into account the different proportions (2/10 vs 2/3).

This situation can be easily solved by introducing a correction factor, $r=\frac{n_+}{n_T}$, defined as the ratio between the number of positive rates $n_+$ and the number of total rates $n_T$.
So we define a new quantifier
\begin{equation}\label{our quantifier with correction}
\mathcal{M}_C=\frac{1}{1+r\sum_{i\,\left(if\,\sigma_i>0\right)}\sigma_i}.
\end{equation}
$\mathcal{M}_C$ looks a better measure of the memory in musical structures than $\mathcal{M}$.

\begin{figure}
\begin{center}
\includegraphics[width=7 cm]{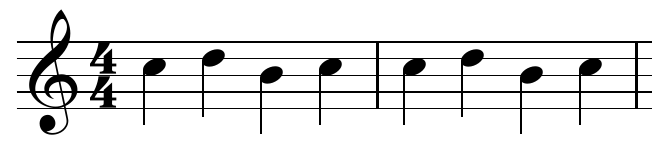} 
\end{center}
\caption{\label{limit_max} \footnotesize Trivial succession of two identical measure:  for the parameters frequency-start, the memory is 1, the maximum value.}
\end{figure}

\begin{figure}
\begin{center}
\includegraphics[width=7 cm]{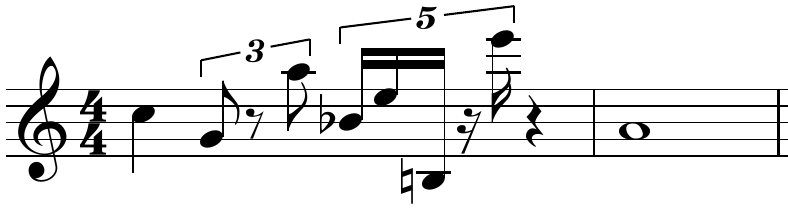} 
\end{center}
\caption{\label{limit_min} \footnotesize The first measure is a quasi-random sequence of pitches and durations, while the second one is totally different: in this case the memory for the parameters frequency-start is 0.3, a very low value, since $0\leq\mathcal{M}_C\leq1$.}
\end{figure}

Now we are giving two simple musical examples. The first one is formed by two identical measures (fig. \ref{limit_max}), and the second one (fig. \ref{limit_min}) presents two uncorrelated measures. The expectation is the maximum value of memory for the first example, and a very lower value for the second one.
Let us consider the parameters frequency and start, and let us evaluate the amount of memory using $\mathcal{M}_C$. In the first example, with two identical measures (fig. \ref{limit_max}), the trace distance (defined in eq. \ref{distance})  between the matrices for the first and second measure is zero, and then the amount of memory is immediately maximal, i.e. $\mathcal{M}_C=1$. However, in the second example, we have two completely uncorrelated measures: the first measure contains quasi-random pitches and duration-start, while the second one has only a whole note. In this case, the amount of memory $\mathcal{M}_C$ is equal to 0.3, a very low value since $0\leq\mathcal{M}_C\leq1$. We have utilized $\mathcal{M}_C$, although in these two examined examples there are only two measures, then there is only one rate $\sigma$. Thus, in this particular case, the results given by $\mathcal{M}$ and $\mathcal{M}_C$ are equivalent.

\maketitle

\section{Application of non-Markovianity criterion \label{examples}}

We apply the methods of the previous chapters, 
and the technique described in chapter \ref{realization} to some existing compositions.

The chosen compositions are the vocal part of the song \textit{Dolente Immagine} that is the 8th piece from \textit{15 Composizioni da Camera} by Vincenzo Bellini \cite{bellini}; the first piece from the oboe suite \textit{Solo} by Bruno Maderna \cite{maderna}, and the piece no.3 from \textit{Metamorphosis I-V} by Philip Glass \cite{glass}.
They are examples of a classic vocal Italian composition, an avant-garde Italian composition and a contemporary minimalist one, respectively.
 
These compositions have been chosen because they appear to have, by ear, rather different degree of musical memory.

The compositions will be analyzed using both the Musical non-Markovianity degree $\mathcal{M}$ given in eq. \ref{our quantifier without correction}, and $\mathcal{M}_C$ of eq. \ref{our quantifier with correction}. We will see that the use of $\mathcal{M}_C$ allows a better characterization of memory in some of the treated cases.

An important step is the division of the structure into sequences.
We will construct tridimensional graphs to represent the structure of pitches, onsets and intensities of musical compositions. Tridimensional graphs to study characteristics of sounds have been proposed by I. Xenakis  \cite{xenakis}, and the use to graphically study orchestration of musical scores by M. Betta \cite{mannone_libro}.

Here we will briefly examine the structure of the chosen compositions, in order to find the optimal subdivision into sequences.

\textit{Bellini}. The vocal score has a structure of type A - B - A', and can be divided into seven periods: sections A has three periods, section B only one, and section A' three ones. The motivation of the subdivision into three sections is due to the tonality change in period 4, and the reprise of theme and of its tonality in period 5. The subdivision into sequences as discussed is due to reasons of musical analysis of a classic model.
There are some identical parts between sections A and A', and this fact induces to expect a high degree of memory.

A matrix is then associated (for each couple of parameters, as discussed in chapter \ref{realization}) to each sequence.
In the case of Bellini a sequence naturally corresponds to a period.

\begin{figure}
\begin{center}
\includegraphics[width=9 cm]{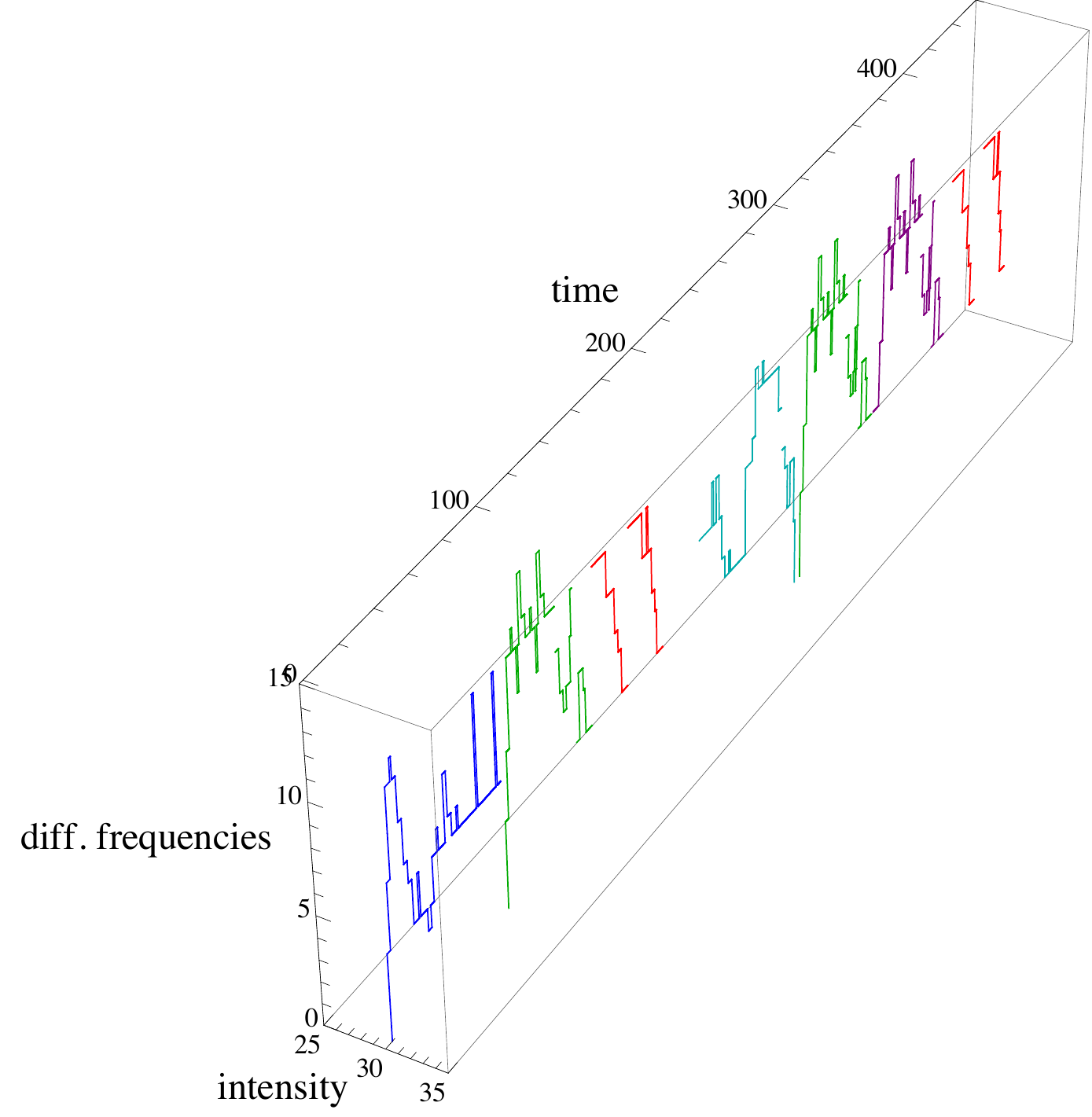} 
\end{center}
\caption{\label{bellini_graph} \footnotesize (Color online) The tridimensional graph represents the vocal part of \textit{Dolente Immagine} by Vincenzo Bellini. 
The image is flat since the intensity is constant.
There are seven periods, and the analyzed sequences correspond to periods. At the same color correspond periods with identical colors. The mean value of $\mathcal{M}_C$ is 0.9.}
\end{figure}

\begin{figure}
\begin{center}
\includegraphics[width=8 cm]{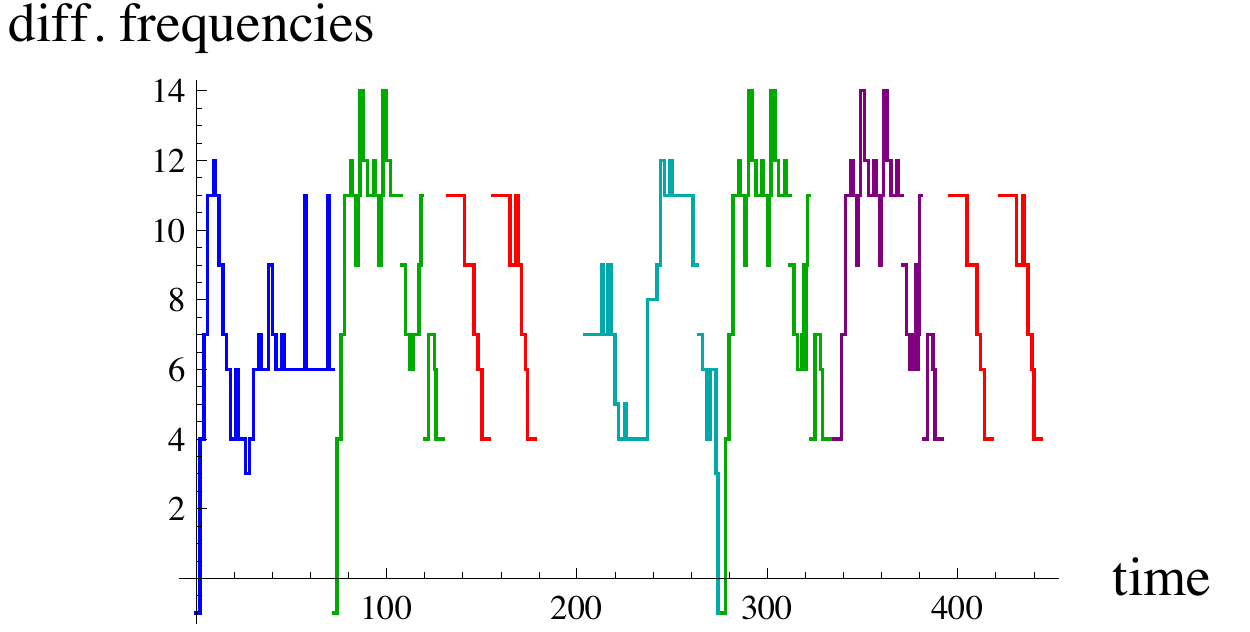} 
\end{center}
\caption{\label{bellini_graph_2d} \footnotesize (Color online) The projection of the graph of fig. \ref{bellini_graph} in the plane time-frequency. The intensity written in the score is constantly equal to $\mathbf{p}$ (\textit{piano}, equal to 30 in our scale), and then the plane time-frequency contains the entire develop of the vocal score. It is evident the presence of repeated patterns.}
\end{figure}

Fig. \ref{bellini_graph} graphically represents pitches, onsets and intensities of the score, showing some repeated patterns. The intensity in the score is constant, and thus the complete development of the composition can be represented in time-frequency plane (fig. \ref{bellini_graph_2d}). 

\begin{figure}
\begin{center}
\includegraphics[width=8 cm]{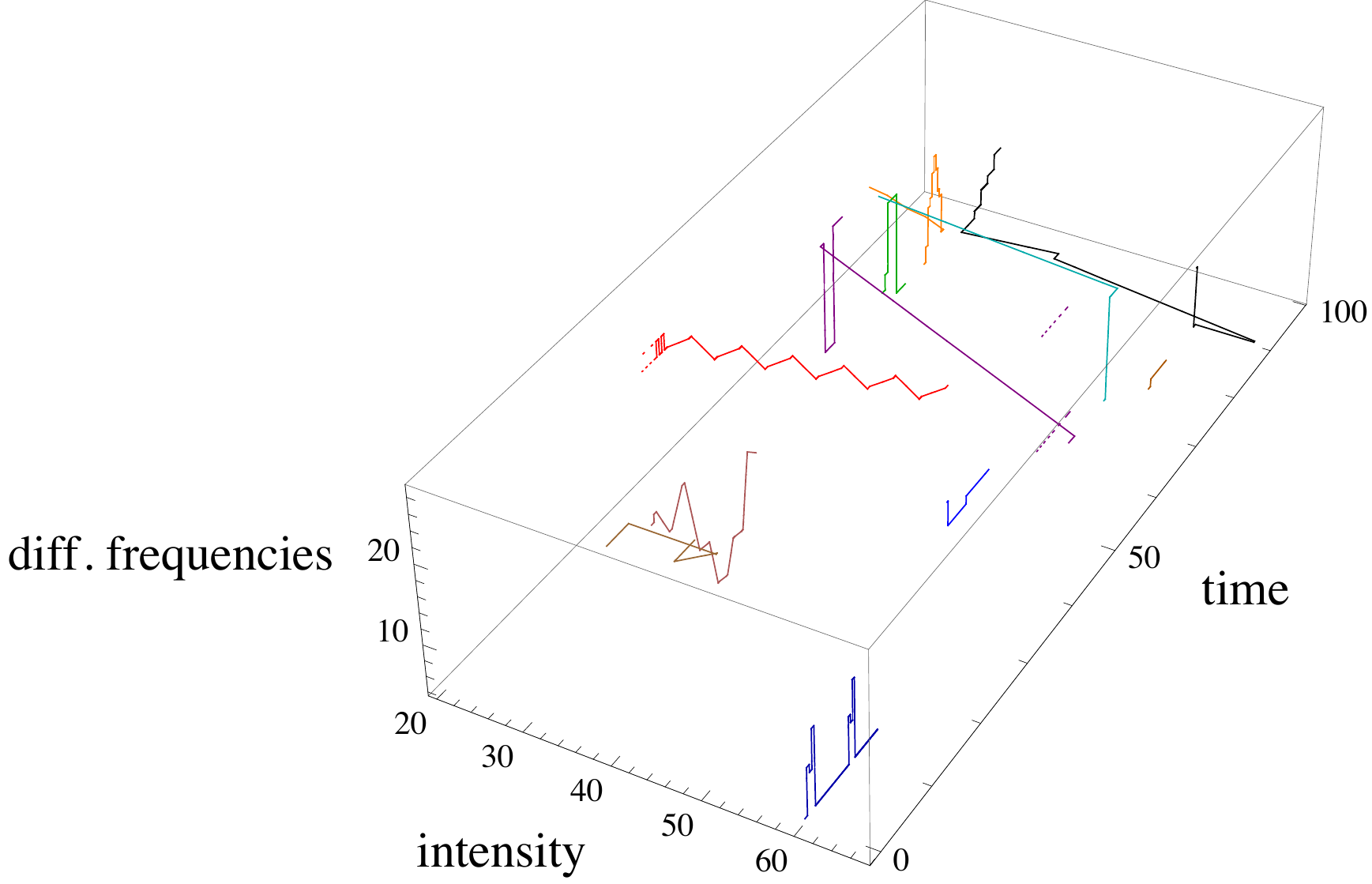} 
\end{center}
\caption{\label{maderna_graph} \footnotesize (Color online) Tridimensional representation of the first piece of the suite \textit{Solo} by Bruno Maderna. It is evident the little quantity of repeated patterns. Each color identifies a different sequence. The mean value of $\mathcal{M}_C$ is 0.6.}
\end{figure}

\begin{figure}
\begin{center}
\includegraphics[width=9 cm]{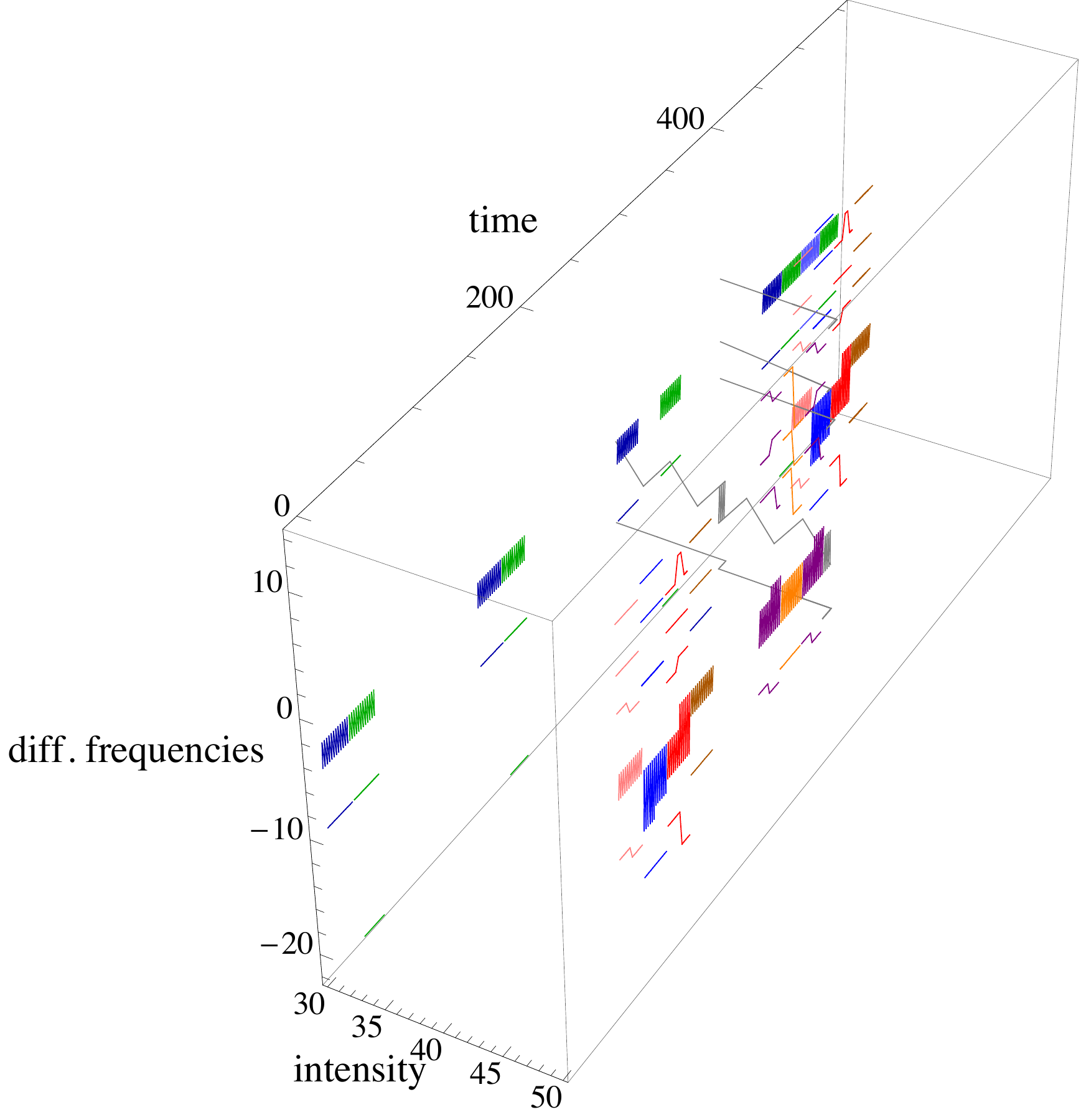} 
\end{center}
\caption{\label{glass_graph} \footnotesize (Color online) Tridimensional representation of the piano composition \textit{Metamorphosis 3} by Philip Glass. In this graph, we chose the value 1 for the eight note (in the other pieces, 2 for the eight note). The calculations of non-Markovianity are unaffected of these variations, because are important distributions towards mean value. The mean value of $\mathcal{M}_C$ is 0.8.}
\end{figure}

\textit{Maderna}. It is immediate by looking at the score of the first piece of the oboe suite \textit{Solo}, and looking at the tridimensional representations of the score in fig. \ref{maderna_graph}, that the structure is very different from Bellini's composition. In fact the quantity of repetitions is clearly lower, there are not thematic or relevant patterns, but only some thematic fragments. So in this case one expects that with respect to the Bellini's score there is a lower degree of memory. Since there were not clearly thematic or tonality motivations as in Bellini's case, it was impossible in this case to talk about periods, but only of sequences separated by the \textit{respiri e legature} (breaths and slurs), the most natural criterion in this case. We have then divided the score into eleven sequences. 

\begin{figure}
\begin{center}
\includegraphics[width=9 cm]{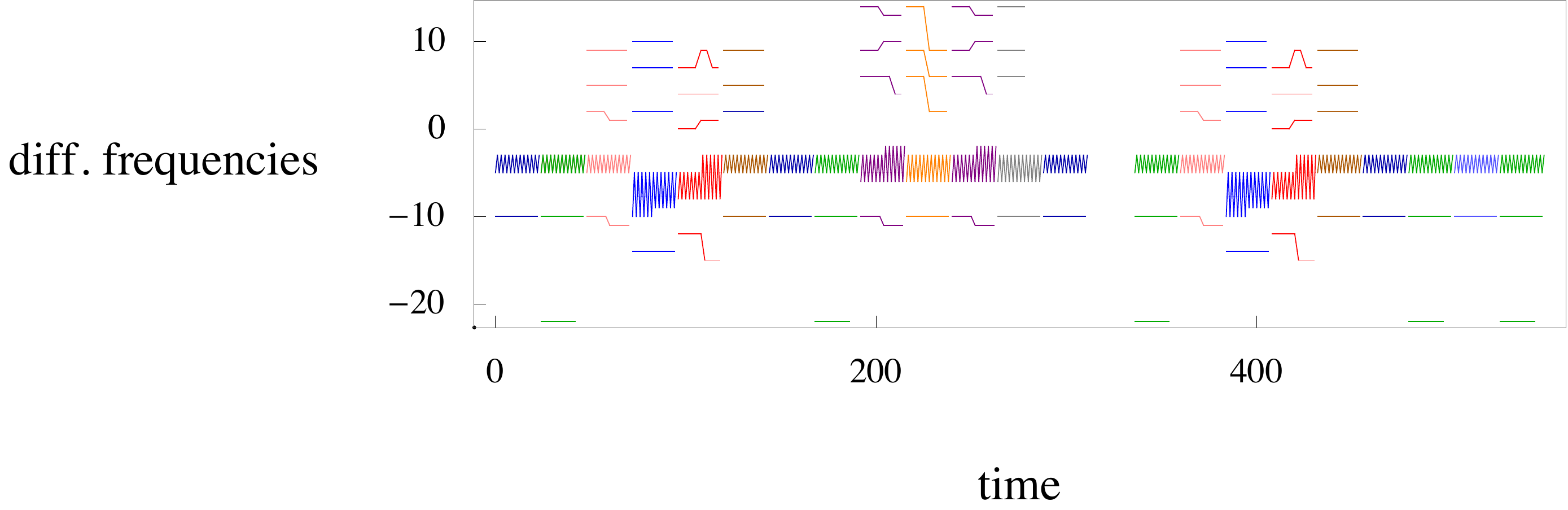}
\end{center}
\caption{\label{glass_graph_2d_1} \footnotesize (Color online) Projection into the plane time-frequency of the graph of fig. \ref{glass_graph}.}
\end{figure}

\textit{Glass}. The piano composition \textit{Metamorphosis 3} is an example of the minimalist style, where there are repetitions with a very small number of pattern changes and variations. The expected degree of memory is therefore higher with respect to the Maderna's case. We have chosen this example because it is a case where it is possible to verify the usefulness of the corrective factor $r$ (eq. \ref{our quantifier with correction}).
Due to its regularity, the composition has been divided into twenty-two sequences, each sequence containing four measures (measure 1 of the refrain has not been considered).
The two-dimensional graph of fig. \ref{glass_graph_2d_1} clearly shows the repeated patterns; the tridimensional graph of fig. \ref{glass_graph} shows also the variations of intensity.

\begin{figure}
\begin{center}
\includegraphics[width=9 cm]{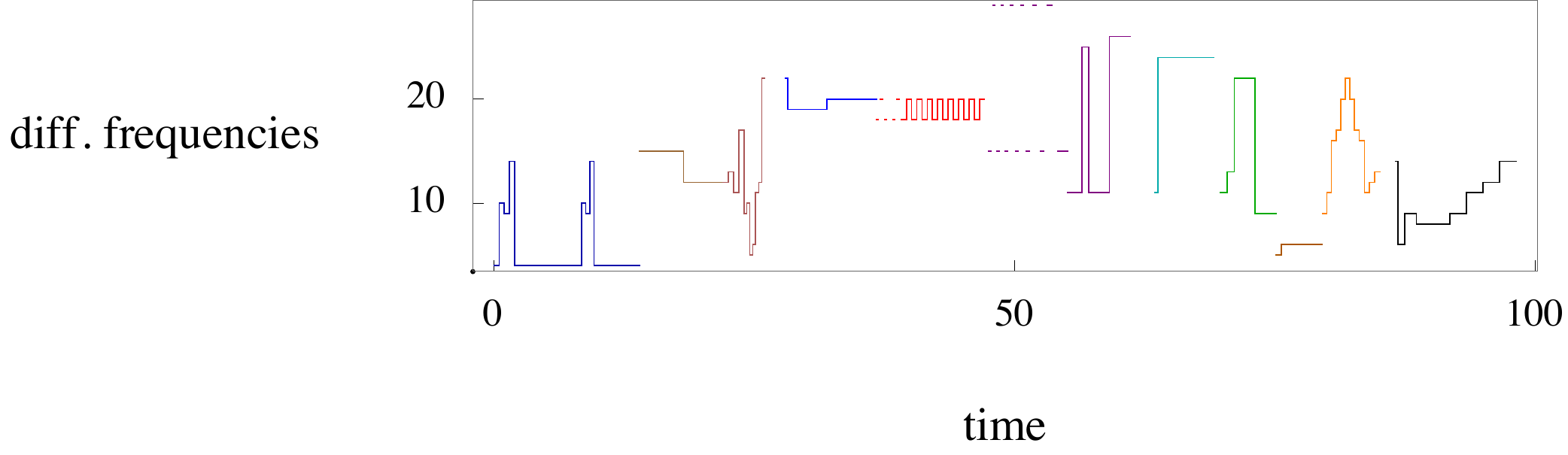}
\end{center}
\caption{\label{maderna_graph_2d_1} \footnotesize (Color online) Projection into the plane time-frequency of the graph of fig. \ref{maderna_graph}.}
\end{figure}

In the following we obtain the musical matrices corresponding to the above musical scores and apply then to analyze the respective memory degree $\mathcal{M}$ and $\mathcal{M}_C$, defined in equations \ref{our quantifier without correction} and \ref{our quantifier with correction}. The results are given in fig. \ref{results}.

\begin{figure}[h]
\begin{center}
\begin{tabular}{|c|c|c|c|}
\hline
\multirow{2}{*}{}& \multicolumn{3}{|c|}{}\\
\multirow{2}{*}{Quantifier}& \multicolumn{3}{|c|}{Results:}\\
\multirow{2}{*}{}& \multicolumn{3}{|c|}{}\\
\multirow{2}{*}{}& \multicolumn{3}{|c|}{}\\
& Bellini &  Maderna & Glass \\
\multirow{2}{*}{}& \multicolumn{3}{|c|}{}\\
\hline
\multirow{2}{*}{}& \multicolumn{3}{|c|}{}\\
\multirow{2}{*}{}&\multicolumn{3}{|c|}{frequency-start}\\
\multirow{2}{*}{}& \multicolumn{3}{|c|}{}\\
\hline
$\mathcal{M}$ & 0.97 & 0.73 & 0.55  \\
\hline
$\mathcal{M}_C$ & 0.99 & 0.77 & 0.97   \\
\hline
\multirow{2}{*}{}& \multicolumn{3}{|c|}{}\\
\multirow{2}{*}{}&\multicolumn{3}{|c|}{duration-intensity}\\
\multirow{2}{*}{}& \multicolumn{3}{|c|}{}\\
\hline
$\mathcal{M}$ & 0.75 & 0.40 & 0.38  \\
\hline
$\mathcal{M}_C$ & 0.82 & 0.53 & 0.78   \\
\hline
\multirow{2}{*}{}& \multicolumn{3}{|c|}{}\\
\multirow{2}{*}{}&\multicolumn{3}{|c|}{start-intensity}\\
\multirow{2}{*}{}& \multicolumn{3}{|c|}{}\\
\hline
$\mathcal{M}$ & 0.86 & 0.54 & 0.51  \\
\hline
$\mathcal{M}_C$ & 0.81 & 0.59 & 0.86   \\
\hline
\multirow{2}{*}{}& \multicolumn{3}{|c|}{}\\
\multirow{2}{*}{}&\multicolumn{3}{|c|}{frequency-duration}\\
\multirow{2}{*}{}& \multicolumn{3}{|c|}{}\\
\hline
$\mathcal{M}$ & 0.75 & 0.56 & 0.55  \\
\hline
$\mathcal{M}_C$ & 0.82 & 0.62 & 0.79   \\
\hline
\multirow{2}{*}{}& \multicolumn{3}{|c|}{}\\
\multirow{2}{*}{}&\multicolumn{3}{|c|}{frequency-intensity}\\
\multirow{2}{*}{}& \multicolumn{3}{|c|}{}\\
\hline
$\mathcal{M}$ & 0.85 & 0.55 & 0.40  \\
\hline
$\mathcal{M}_C$ & 0.94 & 0.67 & 0.57   \\
\hline
\end{tabular}
\end{center}
\caption{\label{results} \footnotesize Values of the Musical non-Markovianity quantifiers $\mathcal{M}$ and $\mathcal{M}_C$ calculated for  musical matrices of the considered examples by Bellini, Maderna and Glass.}
\end{figure}

The values of $\mathcal{M}$ does not appear as correspondent to the empirical expectations for all pieces (although it corresponds for the first two pieces); the values of $\mathcal{M}_C$ appear to correspond: in fact, Bellini's and Glass' compositions have an average degree of memory higher than Maderna's one, and the one of Glass' composition is higher than Bellini's one (using the corrective factor).
In particular, the mean values of $\mathcal{M}_C$ over the considered couples of parameters are: 0.9 for Bellini, 0.6 for Maderna and 0.8 for Glass. 
Therefore, we have seen that the degree of memory $\mathcal{M}_C$ is better than $\mathcal{M}$, since it quantifies correctly the memory of case of Glass score. In some cases the values of $\mathcal{M}$ and $\mathcal{M}_C$ are not significantly different, but in other cases, for example depending on the length of the composition, the correction appears decisive.
Therefore, it seems that the method can be extended to analyze the degree of memory of musical compositions using a larger set of data.

To apply this methods to analyze musical scores it would be advantageous to automatize it to obtain  the reduction into numbers of the musical parameters contained in the score and the subdivision into sequences \cite{mannone_IRCAM}.

\maketitle

\section{Conclusions \label{conclusions}}

We have proposed a method to measure the amount of memory in a musical composition. To quantify musical memory we have constructed quantifiers following a non-Markovianity quantifier used in the theory of open quantum systems \cite{breuer2009}.
For such a purpose we have transformed musical scores into numerical matrices relative to couples of parameters.

In order to obtain matrices, we have firstly divided each musical score into sequences. Chosen a couple of parameter, e.g. pitch and time of start (onset), we have evaluated the mean values of all values of pitch and of all values of onset in the sequence. We have then evaluated the distance (normalized between 0 and 1) of the parameters (a value of pitch and a value of onset) associated to each note in the examined sequence respect to the corresponding mean value (mean pitch and mean onset). Counting the number of notes with a pitch-value in a given interval of distance respect the mean value of pitches and an onset-value in a given interval respect the mean value of onsets, we have constructed the matrix pitch-onset for the considered sequence. We have followed the same procedure for all sequences, and for the various couples of parameters. 

We have then evaluated the trace distance between matrices relative to the same couple of parameters.
The rate of variation of the trace distance between matrices gives information about the memory in open quantum systems, where matrices represent different quantum states.
Since we have normalized musical matrices on the number of notes of each sequence, the evaluation of the rates $\sigma$ is simply the evaluation of differences between matrices. Positive values of $\sigma$ represent increasing distances.

The fundamental difference between measuring memory in physics and in music is that in physics the higher is the distance between initial equal states, the higher is the memory, where in music the higher is the distance, the lower is the musical similarity between sequences and then the lower is the musical memory.

To check the consistence of our quantifier, $\mathcal{M}_C$, that is if the amount of memory measured, we have applied this method to two simple cases: two equal measures in the first example, two very different measures in the second one.
We have found that, in these simple cases, there is agreement between the information given by the quantifier and the memory as defined by ear.

We have applied our method to evaluate the amount of memory of three compositions, that, by ear, show different quantity of memory. These compositions are: The vocal part of the song \textit{Dolente Immagine} by Vincenzo Bellini \cite{bellini}, the first piece from the oboe suite \textit{Solo} by Bruno Maderna \cite{maderna}, the \textit{Metamorphosis 3} by Philip Glass. Qualitatively, the amount of memory of the examined scores by Glass (an example of minimalism) and by Bellini (a score with the classic structure A-B-A) are higher than the memory of Maderna's score (an example of avant-garde).   

The values obtained of $\mathcal{M}_C$ correspond to the expectations. That is $\mathcal{M}_C$ has a higher value in scores just where one, by ear, attributes a larger amount of memory.

The proposed method is a further tool that can be utilized by musicians and musicologists to connect mathematical/physical to musical concepts. It could for example be used to see if different composers in various historical periods can be characterized by a given mean memory level of their compositions and also to compare two similar pieces \cite{Kramer}.
An average value of musical memory could for example characterize the works of the same composer, or of all the composer in the same artistic movement, or a given musical form (\textit{fugue}, \textit{theme with variations}, \textit{fantasia}, and so on).
This technique could be automatized in each step, to allow easy comparisons between larger musical compositions.

\maketitle

\section{Acknowledgements}

We are grateful  to the composer Marco Betta, for fruitful discussions about the concept of memory in music.

\maketitle

\section{Appendix \label{appendix}}

To obtain musical matrices, we have developed an algorithm, with the following steps:
1. evaluation of the normalized distance, for each coordinate, from its mean value;
 2. count of the number of notes in a given interval (from the mean value) e.g. between 0 and 0.25, 0.25 and 0.5, 0.5 and 0.75, 0.75 and 1.

We choose, as an example, only the couple of parameters frequency ($\nu$) vs time of start ($\tau$) of each note in the same sequence.
We consider the following short sequence of sounds as an example.
\begin{center}
\includegraphics[width=3 cm]{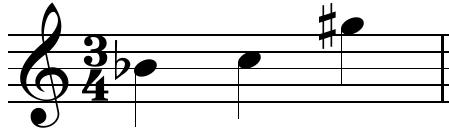}
\end{center}
The musical parameters are frequencies, durations, starts, intensities.
In this example for simplicity there is not any intensity indication.
Let we consider only frequencies and starts.
The frequencies (indicated as the semitones distances from middle C) are $
\nu_1=10,\,\nu_2=12,\,\nu_3=20$,
and the times of start are (the durations are all equal, and corresponding to a crotchet, i.e. a quarter note, = 4 in our scale):
$\tau_1=0,\,\tau_2=4,\,\nu_3=8$.
The distances from the mean values are
\begin{equation*}
 \delta \nu_1=4,\,\,\delta \nu_2=2,\,\,\delta \nu_3=6;\,\,\, \delta \tau_1=4,\,\,\delta \tau_2=0,\,\,\delta \tau_3=4.
\end{equation*}
To normalize these distances, the algorithm find minimum and maximum distance values:
\begin{equation*}
\delta\nu_{min}=2,\,   \delta\nu_{max}=6;\,\,\delta\tau_{min}=0,\, \delta\tau_{max}=4.
\end{equation*}
The normalized distances are evaluated as (if $\delta_{max}$ and $\delta_{min}$ are equal, the algorithm does not normalize):
\begin{equation*}
\delta\nu_i^N=\frac{\delta \nu_i-\delta\nu_{min}}{\delta \nu_{max}-\delta\nu_{min}},\,\,\,
\delta\tau_i^N=\frac{\delta \tau_i-\delta\tau_{min}}{\delta \tau_{max}-\delta\tau_{min}},
\end{equation*}
where i = 1, 2, 3 (in the example considered there are only three notes).
In our example we obtain
\begin{equation*}
\begin{split}
& \delta\nu_1^N=0.5,\,\,  \delta\nu_2^N=0,\,\, \delta\nu_3^N=1,
\\& \delta\tau_1^N=1,\,\, \delta\tau_2^N=0,\,\, \delta\tau_3^N=1.
\end{split}
\end{equation*}
Now it is possible construct the matrix frequency-start for this sequence.
The matrix will contain, in the rows, the number of notes with normalized distance between 0 and 0.25, 0.25 and 0.5, 0.5 and 0.75, 0.75 and 1 from the mean value of start; and, in the columns, the number of notes with normalized distance between 0 and 0.25, 0.25 and 0.5, 0.5 and 0.75, 0.75 and 1 from the mean value of frequency.


To avoid the difficulty of the different length of each sequence (each sequence can contain a different number of notes, the notes can have different durations, the sequence in total can have different durations...), we divide each matrix element by the total number of notes in the considered sequence. In this way, we obtain a  matrix, and the sum of each element is 1. So we can then easily compare sequences of different length, and with different numbers of notes.
In our simple example, the matrix is
\begin{equation*}
\rho=
\left(
\begin{matrix}
0.33 & 0 & 0 & 0
\\ 0 & 0 & 0 & 0
\\ 0 & 0 & 0 & 0.33
\\ 0 & 0 & 0 & 0.33
\end{matrix}
\right)
\end{equation*}



\bibliography{references}

\end{document}